\documentclass[preprint,12pt]{elsarticle}
\usepackage{algorithmic}
\usepackage{graphics}
\usepackage{multirow}
\usepackage{booktabs}
\usepackage{threeparttable}
\usepackage{color,colortbl}
\usepackage{amsmath}
\usepackage{amssymb}
\usepackage{hyperref}
\usepackage{makecell}
\usepackage{mycustomstyle}
\newcommand{\R}[1]{\textcolor{red}{\textbf{#1}}}
\makeatletter
\def\ps@pprintTitle{%
   \let\@oddhead\@empty
   \let\@evenhead\@empty
   }
\makeatother

\begin{document}

\begin{frontmatter}



\title{Identifying vital nodes through augmented random walks on higher-order networks}


\author[label1,label2]{Yujie Zeng\corref{cor1}}

\author[label1,label2]{Yiming Huang\corref{cor1}}
\cortext[cor1]{Equal Contribution}
\cortext[cor2]{Corresponding author}

\author[label1,label2]{Xiao-Long Ren\corref{cor2}}
\ead{renxiaolong@csj.uestc.edu.cn}

\author[label3,label1,label2]{Linyuan L\"u\corref{cor2}}
\ead{linyuan.lv@uestc.edu.cn}

\address[label1]{Institute of Fundamental and Frontier Studies, University of Electronic Science and Technology of China, Chengdu 611731, PR China}
\address[label2]{Yangtze Delta Region Institute (Huzhou), University of Electronic Science and Technology of China, Huzhou 313001, PR China}
\address[label3]{School of Cyber Science and Technology, University of Science and Technology of China, Hefei 230026, China}


\begin{abstract}
Empirical networks possess considerable heterogeneity of node connections, resulting in a small portion of nodes playing crucial roles in network structure and function. Yet, how to characterize nodes’ influence and identify vital nodes is by far still unclear in the study of networks with higher-order interactions. In this paper, we introduce a multi-order graph obtained by incorporating the higher-order bipartite graph and the classical pairwise graph, and propose a Higher-order Augmented Random Walk (HoRW) model through random walking on it. This representation preserves as much information about the higher-interacting network as possible. The results indicate that the proposed method effectively addresses the localization problem of certain classical centralities. In contrast to random walks along pairwise interactions only, performing more walks along higher-order interactions assists in not only identifying the most important nodes but also distinguishing nodes that ranked in the middle and bottom. Our method outperforms classical centralities in identifying vital nodes and can scale to various tasks in networks, including information spread maximization and network dismantling problems. The proposed higher-order representation and the random walk model provide novel insights and potent tools for studying higher-order mechanisms and functionality.
\end{abstract}
 
\begin{keyword}
Vital Node Identification \sep
Node Centrality \sep 
Random Walks \sep
Higher-order Networks \sep
Complex Networks
\end{keyword}

\end{frontmatter}


\section{Introduction}
Over the past two decades, the research on complex networks has undergone remarkable advancements, engendering fruitful applications across diverse domains, such as sociology, biology, and technology, in networked systems \cite{barabasi2016network,newman2018networks}.
Most previous studies in network science primarily concentrate on systems with pairwise topological structures, modeling them as networks comprising sets of nodes and pairwise \add{edges} \del{links}.

\add{However, many real-world networks exhibit non-trivial higher-order structures that cannot be sufficiently analyzed using traditional pairwise methods. This includes networks with community structures, nested hierarchies, and overlapping memberships.
In functional aspect, many networked systems' functionality is influenced or determined by higher-order interactions, wherein such interactions occur not only between node pairs but also involve larger assemblies of nodes simultaneously \cite{battiston2020networks}. 
Instances of these higher-order group dynamics include product recommendations from multiple friends within social networks \cite{SocialNet2010}, information transmission in brain networks necessitating the coordinated activity of numerous neurons \cite{BrainNet2011}, and hunting behaviors involving multiple organisms in ecological competitive networks \cite{Ecology2017}.
Such higher-order group dynamics are a crucial part of empirical systems, but classical pairwise structures are insufficient to capture or simulate these higher-order group dynamics precisely.
Moreover, existing methods might not effectively discriminate nodes in cases where nodes share similar pairwise connectivity but exhibit distinct roles within higher-order structures. Higher-order methods can offer better discrimination between nodes.}
Therefore, the study of higher-order interactions in complex systems is increasingly in the limelight of scientific attention recently \cite{battiston2021physics}.

Nodes are fundamental elements of networks, and it is widely recognized that the uneven influence of nodes in both pairwise and higher-order networks stems from the heterogeneity of network connections \cite{battiston2020networks, albert2000error}. Consequently, different nodes serve quite varied functions, making the determination of node influence and identification of vital nodes the long-term hot topics within the realm of complex networks \cite{lu2016vital}.

In pairwise networks, three primary categories of methods are employed to identify vital nodes: iterative refinement centralities, neighborhood-based centralities, and path-based centralities \cite{lu2016vital}. 
\add{
Iterative refinement centralities consider the support a node receives from its neighbors. Specifically, a node's importance is determined not only by the number of its neighbors but also by the strength of its neighbors. 
Classical methods in this category include Eigenvector centrality \cite{bonacich1987power}, Quasi-Laplacian \cite{Quasi-Laplacian}, PageRank \cite{brin1998anatomy} which is a special kind of random walk.
The random walk method is a fundamental dynamical process in the networks and is highly effective for both local and global network exploration \cite{masuda2017random}. It enables nodes to navigate the network randomly, effectively exploring various aspects of the network's structure. This aids in disseminating and acquiring information in node ranking problems. Additionally, it is not influenced by a specific starting node or path in node ordering, ensuring fair and comprehensive ranking results without any inherent bias. Random walk finds applications in a broad spectrum of complex networks, encompassing social networks \cite{SocialNet2010}, network recommendations \cite{SocialNet2010}, among others, establishing it as a versatile ranking technique.
Nevertheless, node ranking methods based on random walks are subject to severe limitations, such as localization \cite{martin2014localization}. 
Specifically, as iterations progress, most resources concentrate on only a few nodes (i.e., the centrality scores of a few nodes will soar), which will significantly reduce the resolution of other nodes and render their differentiation much more difficult. 
Moreover, classical random walk models defined on pairwise structures inherently fail to capture the characteristics of higher-order structures \cite{battiston2020networks}, thus overlooking the integrity of the whole network.
As for neighborhood-based centralities, such as degree \cite{maharani2014degree}, coreness \cite{kitsak2010identification}, leverage centrality \cite{leverage}, and clustered-local-degree (CLD) \cite{CLD}, are vital for comprehending local network organization and functionality.}
In terms of path-based centrality, node importance is measured by capturing path characteristics between nodes from a propagation perspective. 
Two well-known metrics in this category are Betweenness \cite{freeman1977set} and Closeness centrality \cite{sabidussi1966centrality}. 
In addition to the above three methods for vital nodes identification, other algorithms include graph neural network models \cite{kumar2022influence}, the entanglement models \cite{Qu2020ANC, Arsham2020}, and the random walk-based gravity models \cite{zhao2022random, curado2023novel, gravityModel2019}.

\add{
Higher-order structures play a key role in shaping various aspects of network dynamics due to the consideration of complex interactions between nodes \cite{battiston2020networks}. 
These structures offer novel insights into information dissemination, epidemic contagion, and network robustness analysis. Nevertheless,  relatively fewer metrics have been proposed to measure node importance in higher-order networks due to their complicated structures. 
Some higher-order random walk models have been introduced by simply extending methods on pairwise networks \cite{mukherjee2016random,kaufman2017high}; however, these methods only allow nodes to walk on adjacent-order simplices \cite{mukherjee2016random,schaub2020random}, making them lack flexibility and hard to extend.
}
Therefore, there is a pressing need to develop novel random walk models and node ranking methods that account for \del{the complexities of} higher-order structures in arbitrary orders, in order to better understand the underlying dynamics and provide a more comprehensive perspective on network behavior.

In this paper, we propose a multiscale node ranking method that exploits a Higher-order Augmented Random Walk (HoRW) model and considers the higher-order structures of the network.
Specifically, we first construct a bipartite graph to represent interactions between the higher-order structures and conventional nodes. 
Then, we obtain a multi-order graph by combining the bipartite graph and the classical pairwise network, where a tuning parameter is employed to coordinate the probability of walks going along pairwise or higher-order interactions. 
The proposed Higher-order Augmented Random Walk (HoRW) process allows random walkers to traverse the multi-order graph, and the stationary random walk probability distribution offers a more comprehensive understanding of node importance.
The proposed approach provides a novel node ranking method at multiple scales, enabling the adjustment in accordance with the higher-order interaction strength in the network.
Moreover, the proposed approach outperforms existing methods in identifying vital nodes as well as exhibiting a high resolution in distinguishing middle and bottom nodes, enabling the differentiation of each node in the network based on its importance, rather than solely identifying the top nodes.
Additionally, the proposed approach has broad applicability across various network-related tasks like link prediction, influence maximization, and vital simplex identification.
In summary, our contributions are three-fold as follows:
\begin{itemize}
   \item We introduce a novel representation — multi-order graphs — for complex systems from a higher-order perspective, together with a Higher-order Augmented Random Walk (HoRW) model.
   \item We present a novel node ranking method based on HoRW that allows multiscale analysis according to the strength of higher-order effects.
   \item We demonstrate HoRW's effectiveness in vital node identification and addressing the localization problem, along with significant performance gains in epidemic spreading and network dismantling experiments.
\end{itemize}

The remainder of this paper is organized as follows. 
Section \ref{sec:Preliminaries} overviews the background knowledge and introduces basic network descriptors, contagion models, and random walk models.
In Section \ref{sec:model-HoRW}, we propose a node ranking method based on augmented random walk dynamics on the higher-order networks and analyze the convergence of this method.
In Section \ref{sec:exp}, we conduct several empirical experiments to evaluate our method's performance in two areas: epidemic spreading and network dismantling. 
To further validate our approach, we also assess its resolution - the ability to distinguish nodes of various importance. 
Finally, Section \ref{sec:conclusion} concludes the paper and brieﬂy explores future research directions.

\section{Preliminaries} 
\label{sec:Preliminaries}

This section introduces the basic concepts and notions of pairwise and higher-order networks, with some existing algorithms to identify vital nodes in networks.

\subsection{Network Descriptors}

Networks (or Graphs) provide a powerful framework for elucidating complex systems, and mathematically, they can be represented as a tuple $G=(V, E)$, where $V$ denotes a collection of nodes (or vertices) and $E$ represents a set of edges connecting nodes in $V$.
Different nodes play significantly different roles owing to the ubiquitous existence of the heterogeneous connectivities of complex networks.
Various centrality metrics are usually used to quantify the importance of nodes from different perspectives. Table \ref{index_intro} summarizes five classic centrality measures and their calculation methods. 
\add{Further elaboration on cutting-edge techniques is deferred to Supplementary Information Section 3.}
However, research on the problem of node ranking in higher-order networks has only recently commenced. We will use the bipartite graph to represent higher-order networks more clearly, and then study the problem of node ranking in higher-order networks. 
We proceed to introduce the concepts and notions of bipartite graphs and simplicial complexes, a typical representation of higher-order networks.

\begin{table}[!ht]
\centering
\caption{Formula and explanation of classic centrality measures.}
\renewcommand{\arraystretch}{1.3} 
\resizebox{\textwidth}{!}{
\begin{tabular}{lp{3.4cm}p{15cm}}
\toprule
\textbf{Centrality} &\textbf{Formula} &\textbf{Explanation}\\
\midrule
\textbf{Degree} &$k_i=\sum_{j \in V}a_{ij}.$ & The number of neighbors connecting to node $i$, where $V$ denotes the set of nodes and $a_{ij}$ is the element $ij$ of the adjacency matrix $A$.  \\

\textbf{Closeness} &$C_i=\frac{1}{\sum_{j \neq i}\ell_{ij}}.$ &The inverse of the sum of the length of the shortest paths, where $\ell_{ij}$ is the shortest path distance between node $i$ and $j$. \\

\textbf{Betweenness} &$B_i = \sum_{i \neq j \neq t} \frac{\sigma_{jt}(i)}{\sigma_{jt}}.$ &The number of shortest paths pass through node $i$, where $\sigma_{jt}$ is the number of shortest paths between nodes $j,t$, and $\sigma_{jt}(i)$  denotes the number of shortest paths between nodes $j,t$ passing through $i$. \\


\textbf{Eigenvector} &$E_i=\frac{1}{\lambda}\sum_{t \in \mathcal{N}(i)}E_t.$ &Eigenvector centrality is calculated by the eigenvector associated with the largest eigenvalue $\lambda$ of the network's adjacency matrix, where $\mathcal{N}(i)$ denotes the neighbors of node $i$. \\

\textbf{PageRank} & $ PR_i^t=\frac{1-a}{n} + a \sum_{j \in \mathcal{N}(i)} \frac{PR_j^{t-1}}{k_j}.$ 
& PageRank is calculated by iteration based on both the quantity and the degree of the neighbors to each node, where $n$ denotes the number of nodes, $a$ represents the damping factor, and $t$ denotes an iterative parameter.  \\
\bottomrule
\label{index_intro}
\end{tabular}}
\end{table}

\textbf{Bipartite graphs} \cite{zha2001bipartite}, also \add{referred to as} \del{called} two-model networks, are characterized by two disparate sets of nodes, wherein connections are exclusively permitted between nodes belonging to different sets. 
Mathematically, a bipartite graph can be represented by an incidence matrix $B$. \del{, a logical matrix that shows the relationship between two sets of objects.}  
Suppose \add{that the} bipartite graph $\mathcal{G}_b$ is composed of two distinct node sets $\mathcal{U}$ and $\mathcal{V}$, then $B_{ij}=1$ \add{signifies the existence of} \del{refers to there exists} an edge from node $i\in \mathcal{V}$ to node $j \in \mathcal{U}$. In this article, we focus on the undirected networks only, thus $B_{ij}=B_{ji}$.
\add{Notably, the utility of bipartite graphs has recently been demonstrated in effectively modeling higher-order interactions \cite{ISMnet}.}
In practice, there are plenty of empirical examples, such as \add{bipartite} user-commodity \del{bipartite} graphs \cite{zhou2007bipartite,ren2014avoiding} and \add{bipartite} author-paper \del{bipartite} graphs \cite{zeng2017science}.



\begin{figure*}[!ht]
\centerline{\includegraphics[width=0.8\linewidth]{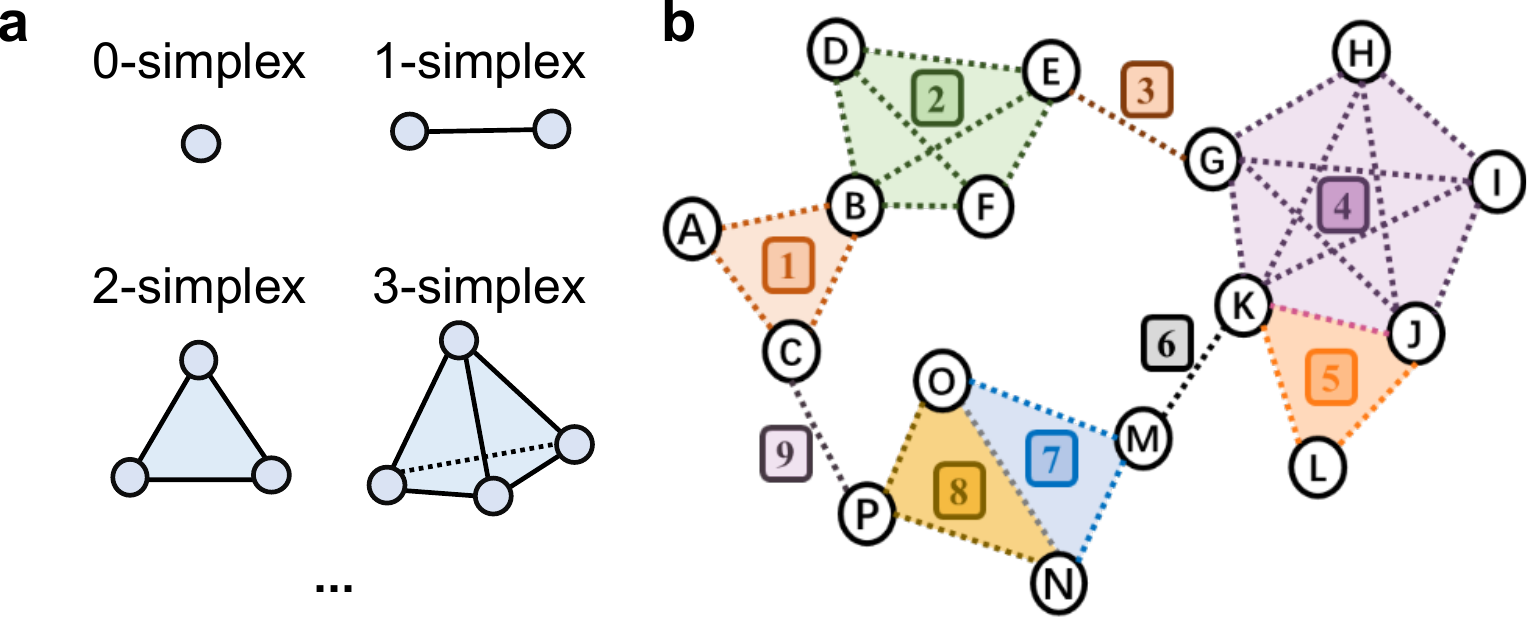}}
\caption{ 
\add{\textbf{Simplicial complex and its matrix representation.}
In \textbf{a}, the illustration shows several prevalent simplices, while their aggregation culminates in the formation of the simplicial complex shown in \textbf{b}. 
}
\del{\textbf{The construction process of the higher-order bipartite graph.} \textbf{a} demonstrates several common simplices. 
\textbf{b} illustrates an example network with simplices in different colors. 
\textbf{c} presents the incidence matrix $B$ for graph \textbf{b}. 
\textbf{d} represents the higher-order bipartite graph derived from the pairwise graph \textbf{b}.}
}
\label{fig:SCs}
\end{figure*}

\textbf{Simplicial complexes (SCs)}
serve as robust mathematical constructs employed to represent topological spaces \cite{battiston2020networks}, facilitating the analysis of their inherent properties and structures through algebraic topology \cite{millan2020explosive}.
In mathematics, a simplicial complex is a collection of simplices of different dimensions closed under taking subsets.
A $d$-simplex is constituted by $d+1$  fully interconnected nodes, encompassing entities such as nodes (0-simplex), edges (1-simplex), ``full'' triangles (2-simplex), and so forth.
\add{See Fig. \ref{fig:SCs} \textbf{a} for a visual illustration.}
Simplicial complexes offer an invaluable framework for characterizing interactions encompassing more than two nodes, transcending the limitations imposed by pairwise structures.

\subsection{Random Walk Models}

Random walk is a stochastic process \cite{masuda2017random} that has found fruitful applications in various fields, such as mathematics, physics, chemistry, and economics.
The dynamics of random walks can be employed to comprehend how information flows are locally trapped in the network and to uncover the intrinsic properties of both nodes and the entire network. It is an effective framework for ranking nodes in networks \cite{newman2005measure}.
Random walks come in many forms \cite{masuda2017random}, including Markov chains, Brownian motion, drunkard's walk, and Lévy flight. 
In practice, the trajectory of a drunkard, the Brownian motion of pollen, and the rise and fall of securities are all inseparably related to random walks.

\textbf{Classical random walk.} 
The main idea of classical random walk is to traverse a graph by starting at a node or a series of nodes and moving to a neighboring node at random. 
\add{Consider} \del{Suppose there is} a graph with \add{ $n$ nodes, $m$ edges,} and an adjacency matrix $A$.
A random walker starts at node $i$ and walks to the neighboring node $j$ with the probability of $A_{ij}/k_j$, \add{indicating}\del{meaning} an equal selection probability between $k_j$ neighbors. 
Let $\pi_i(t)$ encode the probability of node $i$ \add{being} \del{to be}occupied by a random walker at time $t$.
\del{This probability distribution will be used as the input for the next walk, and the process is iterated repeatedly as}
\add{Hence, the random walk process is governed by}
\begin{equation}
    \pi_i(t) = \sum_j \frac{A_{ij}}{k_j} \pi_j(t-1).
\end{equation}

We can equivalently describe the above process as 
\begin{equation}
    \pi(t)=C\pi(t-1),
\end{equation}
where $\pi(t)={\left(\pi_1(t),\cdots,\pi_n(t)\right)}^\top$,  $C=AD^{-1}$ is the transition matrix encoding the transition probability, and $D = \operatorname{diag}\left(k_1,\cdots,k_n\right)$ is the degree matrix. 
\del{The probability that a walker moves from node $i$ to $j$ in a one-time step is based on the $(i,j)$ entry of the transition matrix $C (C_{ij}=A_{ij}/k_j)$.}

In a connected network, the probability of a random walk to node $i$ is positively related to the degree of $i$ after enough steps, i.e., $\pi_i(t\rightarrow \infty)$ will reach convergence \cite{masuda2017random}, and it can be obtained that
$
    \pi_i = {k_i}/{2m}.
$
\del{Here, $k_i$ denotes the degree of node $i$, and $M$ is the number of edges.} An intuitive understanding of the above conclusion is that nodes with greater degrees own more paths to them and hence are more likely to be randomly accessed.

\textbf{PageRank.} 
The PageRank algorithm \cite{brin1998anatomy}, founded upon random walk dynamics within the web graph, \add{is a cornerstone of web search engines like Google.} \del{is widely employed by Google and other search engines to rank web pages.} 
It assigns a quantitative importance score to each page in a graph representing the web, based on the number and quality of incoming links to the page.
Specifically, a hypothetical surfer moves from one web page to another by following hyperlinks, simulating the browsing behavior of real users. 
To account for the possibility that a surfer might get bored and jump to a random page, rather than following a link, PageRank also incorporates a damping factor.
\add{Essentially,} the importance of a web page \add{correlates with} \del{is proportional to} the probability of a random surfer landing on that page after a series of random jumps.

\textbf{Higher-order random walk.} 
Higher-order structures have been successful in providing novel insights into the random walk process \cite{battiston2020networks}.
The random walk on SCs, a type of higher-order random walk, has been extensively investigated \cite{schaub2020random}.
Mukherjee et al. \cite{mukherjee2016random} \add{propose}\del{introduce} a class of random walks on SCs with absorbing states that relate to the spectrum of the $k$-dimensional Laplacian. 
Kaufman et al. \cite{kaufman2017high} introduce the concept of higher-order random walk on high dimensional SCs and develop a local-to-global criterion to ensure the rapid convergence of all higher-order random walks. 
These approaches rely on a novel concept of high-dimensional combinatorial expansion, known as colorful expansion.

In addition to random walks on SCs, other kinds of higher-order random walks include random walks on hypergraphs \cite{Zhou20071601,hayashi2020hypergraph,benson2017spacey}, spacey random walks \cite{benson2017spacey}, and random walks on multiple networks \cite{Luo9946430}.
\del{Hypergraphs are more flexible representations of higher-order networks, and one trivial random walk model in hypergraphs \cite{Zhou20071601} allows the walker to randomly pick one hyperedge containing the current node, and subsequently walk randomly to a node in the selected hyperedge.}
\del{Carletti et al. \cite{carletti2020random} propose a new class of random walks on hypergraphs that can describe systems with multi-body interactions without any limitation to the sizes of the hyperedges. 
They \cite{carletti2021random} further suggest a family of random walk processes on hypergraphs that can be biased towards hyperedges of low or high cardinality by a parameter.}
%
\del{These basic approaches can extend to many existing models of random walks on hypergraphs, \cite{hayashi2020hypergraph}.}
\del{Benson et al. \cite{benson2017spacey} propose the spacey random walks which use a random walk process based on the simplicial complex, where the probability of transitioning from one node to another is proportional to the similarity between the two nodes in the complex.}
\del{Luo et al. \cite{Luo9946430} extend the concept of random walk on multiple networks to include many-to-many node mappings between networks, providing a comprehensive framework for random walk research.}

\subsection{Contagion Models}

Simple contagions can suffice to describe certain social contagion phenomena, such as easily convincing rumors or domino effects \cite{anderson1992infectious}. 
However, they fail to provide a satisfactory explanation in other instances, particularly when more intricate dynamics of peer influence and reinforcement mechanisms are at play \cite{iacopini2019simplicial}.
In the following, we present both the classical SIR model and its higher-order extension, the HSIR model.

\textbf{SIR contagion model.} The SIR (susceptible-infected-recovered) model \cite{lloyd2001viruses} is a classical model for epidemic disease spread.
The whole population can be categorized into three groups: susceptible, infected, and recovered individuals, with each individual. 
Each node in this model can exist in one of these three discrete states.
The susceptible (S) is attributed to people who have\del{n't} \add{not} contracted the infection yet but are at risk of doing so.
The infected state (I) applies to people who have contracted the infection and are capable of transmitting it to others who are susceptible. 
The recovered state (R) refers to people who have already recovered from the infection and are no longer at risk of contracting or transmitting it.
%
%
%
At the beginning of the simulation, all nodes are in the susceptible state, except for a few nodes that are already infected state. 
The infected nodes can spread disease to susceptible nodes with a probability of $\beta$, while they themselves can recover with a probability of $\gamma$, leading to a transition to the recovered state.
This iterative process continues until there are no more infected nodes in the network.

%
\textbf{Higher-order SIR (HSIR) contagion model.}
%
Interactions between pairs of individuals are frequently inadequate to portray complicated social contagion processes. Therefore, it's necessary to construct HSIR contagion model \cite{iacopini2019simplicial}, which employs not only edges as spreading mediums, akin to the standard SIR model, but also incorporates simplices as extra contagion mediums.
Specifically, in the HSIR model, if all nodes in a $D$-simplex are infected except one susceptible node, then the remaining susceptible node is infected by the $D$-simplex and all its lower-dimensional simplices at the respective rates of $\beta_{1}, \beta_{2}, \cdots, \beta_{D-1}$.
Note that, $\beta_1$ is equal to the infection probability $\beta$ in the standard SIR model, wherein a susceptible node $i$ contracts the infection from an infected neighbor $j$ through the \add{edge}\del{link} $\left[i,j\right]$.
Hence, we employ $\beta$ instead of $\beta_1$ to simplify the notation.
Similarly, the second parameter $\beta_2$ corresponds to the probability per unit time that node $i$ receives the infection from a 2-simplex (``full” triangle)  $\left[i,j,k\right]$ where both $j$ and $k$ are infectious, and so on.

As for simulation, the probability of infection for node $i$ is calculated as:
\begin{equation}
P_i=1-(1-\beta)^{n_1} (1-\beta_{2})^{n_2} \cdots
\end{equation}
Here, $n_p$ represents the numbers of $p$-simplex containing node $i$ and whose nodes are all infected except for node $i$.
%



\begin{table}[!ht]
\centering
\caption{Frequently-used symbols in \add{the Methodology section} \del{Section 3}.}
\begin{tabular}{cl}
\toprule
Notations &Description\\
\midrule
$G(V,E)$ &The pairwise graph \\
$\mathcal{G}_b(V,\mathcal{W},\mathcal{E})$ &The bipartite graph\\
$\mathcal{W}=\{\alpha_i\}$ &The simplicial node set of $\mathcal{G}_b$\\
$\pi(t)\in \mathbb{R}^{|V|}$ & The random walk probability distribution at time $t$\\
$A$ &The adjacency matrix of $G$\\
$B$ &The incidence matrix of $\mathcal{G}_b$\\
$C$ &The classical transition matrix\\
$U$ &The upstream transition matrix\\
$D$ &The downstream transition matrix\\
$W$ &The transition matrix on $\mathcal{G}_b$\\
$\widetilde{M}$ &The augmented transition matrix\\
$\Lambda_v$ &The degree matrix of the original nodes in $\mathcal{G}_b$\\
$\Lambda_s$ &The degree matrix of the simplicial nodes in $\mathcal{G}_b$\\
\add{$\varepsilon$} & \add{convergence threshold} \\ 
$s$ &The tuning parameter\\
\bottomrule
\end{tabular}
\label{tab4}
\end{table}

\section{\add{Methodology} \del{Higher-order Augmented Random Walk (HoRW)}}
\label{sec:model-HoRW}
This section is dedicated to delineating the mathematical foundations and methodological procedures employed throughout this article.

\unskip

\subsection{Multi-order Graph}

In this section, we introduce the multi-order graph as a powerful tool for the representation and visualization of higher-order networks. Compared with some classical methods, the multi-order graph can preserve a more complete structure of the higher-order interactions, and give a clearer picture for information delivery.

Classical networks have heretofore been limited to exploiting pairwise interactions between nodes exclusively, yet it has become increasingly evident that higher-order structures significantly influence network dynamics \cite{battiston2020networks}. 
Higher-order structures do not simply arise from the aggregation of pairwise interactions; rather, they embody indivisible entities themselves. 
For example, a 2-simplex conveys information exceeding the contributions of the three individual nodes comprising it \cite{ISMnet,iacopini2019simplicial}. 

%
To fully capture the importance and integrity of these higher-order structures, we introduce a \textbf{higher-order bipartite graph} $\mathcal{G}_b=(V,\mathcal{W},\mathcal{E})$, in which each higher-order structure is treated as an independent node, namely a higher-order structural node. 
Here, $V$ and $\mathcal{W}$ signify the set of original nodes and higher-order structural nodes, respectively, and each higher-order structural node is connected to the constituent nodes that make up the higher-order structure.
Furthermore, we introduce a \textbf{multi-order graph}, which integrates the classical pairwise graph and higher-order bipartite graph with tuning weight $s$.

\begin{figure}[!ht]
\centerline{\includegraphics[width=1.1 \linewidth]{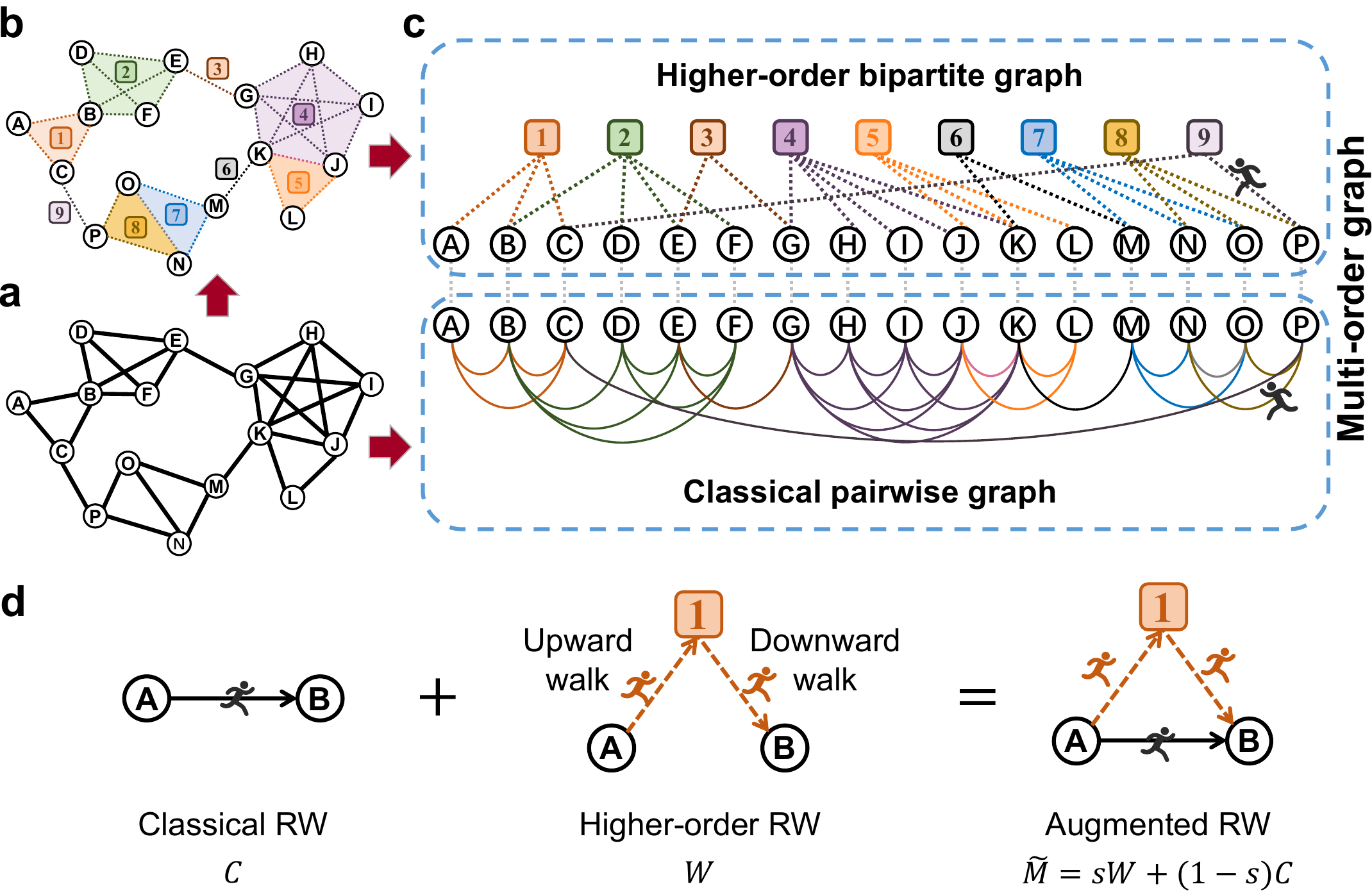}}
\caption{
\add{\textbf{Visualization of multi-order graph construction and augmented random walk process.}
In \textbf{a}, we present a pairwise graph, which can be transferred into the simplicial complex shown in \textbf{b} by identifying the simplices within this graph. 
The upper panel in \textbf{c} illustrates the higher-order bipartite graph derived from the SC in \textbf{b}.
The multi-order graph showcased in \textbf{c} integrates the classical pairwise graph with the obtained higher-order bipartite graph.
\textbf{d} illustrates the augmented random walk (RW) conducted on the introduced multi-order graph, which incorporates the classical and the higher-order random walk with tuning weight $s$. 
Here, $C$ denotes the classical transition matrix based on the pairwise graph, $W$ represents the transition matrix based on the higher-order bipartite graph, 
Here, $C$ and $W$ denote the classical transition matrix on the pairwise graph and the transition matrix on the higher-order bipartite graph, respectively, and $\widetilde{M}$ is the augmented transition matrix capturing the random walk dynamics in the proposed multi-order graph.
}
\del{\textbf{The illustration of multi-order graph and augmented transition matrix $\widetilde{M}$.} 
The multi-order graph integrates the classical pairwise graph and higher-order bipartite graph with tuning weight $s$.
$W$ denotes the transition matrix based on the higher-order bipartite graph, $C$ is the classical transition matrix based on the pairwise graph, and $\widetilde{M}$ is the augmented transition matrix that integrates $W$ and $C$.}
}
\label{fig:framework}
\end{figure}

In this paper, we focus our study on simplicial complexes. 
Thus, to construct a higher-order bipartite graph $\mathcal{G}_b$ based on a pairwise network, the first step involves traversing the entire network to identify all the maximum simplices and enumerate them in succession.
Each node is then connected to the corresponding simplicial node in the higher-order bipartite graph. Consequently, the resulting higher-order bipartite graph $\mathcal{G}_b$ encompasses the higher-order topological features of the original network.
A demonstration of this transformation is given for the toy example in Fig. \ref{fig:framework}.

Drawing inspiration from incidence matrices in pairwise networks, we can analogously define a higher-order incidence matrix $B$ to facilitate investigations of $\mathcal{G}_b$. 
The incidence matrix characterizes the relationship between higher-order structures and nodes, in which columns represent nodes, rows correspond to simplices, and the entities of the matrix indicate their affiliation.
Mathematically, it is defined as:
\begin{equation}
    B_{ij} = \left\{
        \begin{array}{cl}
        1, & j \in {\alpha_i} \\ 
        0, & \text { otherwise } 
        \end{array}\right. .
\end{equation}
Here, $B_{ij}=1$ indicates that node $j$ is contained by simplex $\alpha_i$ (i.e., $j \in {\alpha_i}$).  

The proposed higher-order bipartite graph presentation can preserve the integrity of the higher-order structure as an indivisible entirety for information delivery.
However, it neglects trivial lower-order interactions modeled by pairwise connections.
Therefore, we further introduce a multi-order graph, which incorporates the classical pairwise graph and the higher-order bipartite graph.
An illustrative example is employed to demonstrate the construction process of the multi-order graph, as seen in Fig. \ref{fig:framework}.

\unskip

\subsection{Augmented Random Walk Dynamics}

Upon establishing the multi-order graph derived from a pairwise graph, we can now broaden the random walk on it for a higher-order implementation.
We have introduced the random walk process on pairwise graphs; henceforth, we concentrate on the random walk dynamics on the higher-order network represented by the variant of the bipartite graph introduced above.
The steps for executing a random walk on $\mathcal{G}_b$ entail random walking between the original nodes and the simplicial nodes. This stochastic process is capable of capturing latent higher-order topological properties of the network and thus we name it as Higher-order Augmented Random Walk (HoRW) model. 
Depending on the random walk direction, it can be \add{divided} \del{bifurcated} into two parts: upward walk and downward walk.

\textbf{Upward Walk}.
The upward walk process entails traversing from the original nodes to their corresponding simplicial nodes in the bipartite graph. 
The importance of each node is evenly distributed among the simplices containing it during the upward walk.
Mathematically, the probability of walking from node $j$ to simplicial node $\alpha_i$ can be expressed as follows:
\begin{equation}
    u_{ij}=\frac{B_{ij}}{k_j}.
\end{equation}
Here, $ k_j$ represents the degree of the node $j$ in $\mathcal{G}_b$, and $k_j = \sum_{l} B_{lj}$.

Equivalently, it can be represented in upstream transition matrix form as
\begin{equation}
    U_{m \times n} = \left( u_{ij} \right) =   B \Lambda_v^{-1},
\end{equation}
where $\Lambda_v$ is a diagonal matrix and the diagonal elements are the degrees of the original nodes in $\mathcal{G}_b$.

\textbf{Downward Walk}.
In the context of the bipartite graph $\mathcal{G}_b$ introduced above, the downward walk can be understood as a transfer of information from the simplices back to the nodes. 
In this process, the importance of each simplex is evenly distributed among the nodes that it contains. 
This ensures that each node receives a proportional allotment of the importance of the simplices to which it belongs.
We can similarly define the downstream transition matrix for the downward walk dynamics in $\mathcal{G}_b$ as:
\begin{equation}
    D_{n \times m} = \left( d_{ij} \right) =   B^{\top}\Lambda_s^{-1}.
\end{equation}
Here, $\Lambda_s$ is a diagonal matrix with diagonal elements \add{equal to} \del{being} the dimension of the simplices. The element $d_{ij}$ in $V_{n \times m}$ \add{gives} \del{indicates} the probability of moving from \add{the} simplex $\alpha_j$ to \add{the} node $i$, and thus $d_{ij}=B_{ji}/{k_{\alpha_j}}$, where $k_{\alpha_j}=|\alpha_j|$ presents the dimension of the simplex $\alpha_j$.

The proposed Higher-order Augmented Random Walk constitutes a two-step procedure, combining upward and downward walks within the ``node-simplex" bipartite graph $\mathcal{G}_b$.
The transition matrix $W_n$ integrates this two-step random walk by multiplying the upstream and downstream transition matrix, such that:
\begin{equation}
    W_n = D_{n \times m} \times U_{m \times n}.
\end{equation}
This two-step procedure facilitates the transfer of information from simplices back to the original nodes.

\add{In addition,} the proposed two-step HoRW model exhibits a tendency for walkers to linger within nodes that are involved in a greater number of higher-order structures. 
Likewise, we note that random walks on bipartite-represented higher-order networks capture only the higher-order characteristics of the network, neglecting the pairwise features.
Therefore, it is deemed more appropriate to perform random walks on both the higher-order bipartite graph and the classical pairwise graph, i.e., on the multi-order graph, to gain a comprehensive understanding of the network's features.



\add{To facilitate this exploration,} we introduce an augmented transition matrix $\widetilde{M}$ to represent the random walk process on the \add{introduced} multi-order graph.
Specifically, considering the diverse types of structures, the augmented transition matrix that incorporates higher-order structures necessitates the inclusion of tuning parameter $s$ to accommodate various network structures.
In mathematics, the augmented transition matrices can be defined with flexibility as follows:
\begin{equation}
    \widetilde{M} = sW + (1-s)C,
\end{equation}
where $C$ denotes the classical transition matrix employed in pairwise networks. $s$ is a tuning parameter ranging from 0 to 1, providing a multiscale framework for node ranking. 

This definition allows the random walk process to account for both pairwise interactions (captured by $C$) and higher-order interactions (captured by $W$), while selectively emphasizing one of them based on the value of $s$.
%
%
When higher-order effects dominate the network, a larger value of $s$ is preferred to prioritize random wandering on the higher-order bipartite graph.  
Conversely, when the network is dominated by pairwise interactions, a smaller value of $s$ should be chosen to give more weight to the pairwise subgraph on the multi-order graph.
Under some conditions, when $s=0$, nodes travel exclusively through the pairwise graph; whereas when $s=1$, nodes traverse solely within $\mathcal{G}_b$.
Furthermore, when $s=0.5$, nodes traverse both higher-order bipartite and pairwise subgraphs with equal probability.
Therefore, $\widetilde{M}$ represents the random walk process on the multi-order graph.

\subsection{\add{HoRW Node Ranking on Toy Example} \del{Node Ranking via HoRW on A Toy Example}}
\label{RankingToy}


By employing the augmented transition matrix $\widetilde{M}$, we can capture both the pairwise and higher-order interactions within the network.
To obtain the node importance ranking, we compute the stationary probability distribution of the augmented random walk process on the multi-order graph. 

Specifically, the state transition equation for augmented random walk in our method is as follows:
\begin{equation}
    \pi(t)=\widetilde{M} \pi(t-1).
\end{equation}
Here, \add{$\pi(t)$} \del{$Rw(t)$} denotes the vector whose $i$-th item represents the probability of reaching node $i$ at \add{step} \del{the moment} $t$.
The iteration terminates once the distribution converges, i.e., \add{$\| \pi(t) \|_2 \leq \varepsilon$, ($\varepsilon=0.01$ is adopted in experiments)}, and the stationary probability distribution ranks nodes in order of importance. 
The proof of convergence will be provided in the subsequent section.

\begin{figure}[!ht]
\centerline{\includegraphics[width=0.8\linewidth]{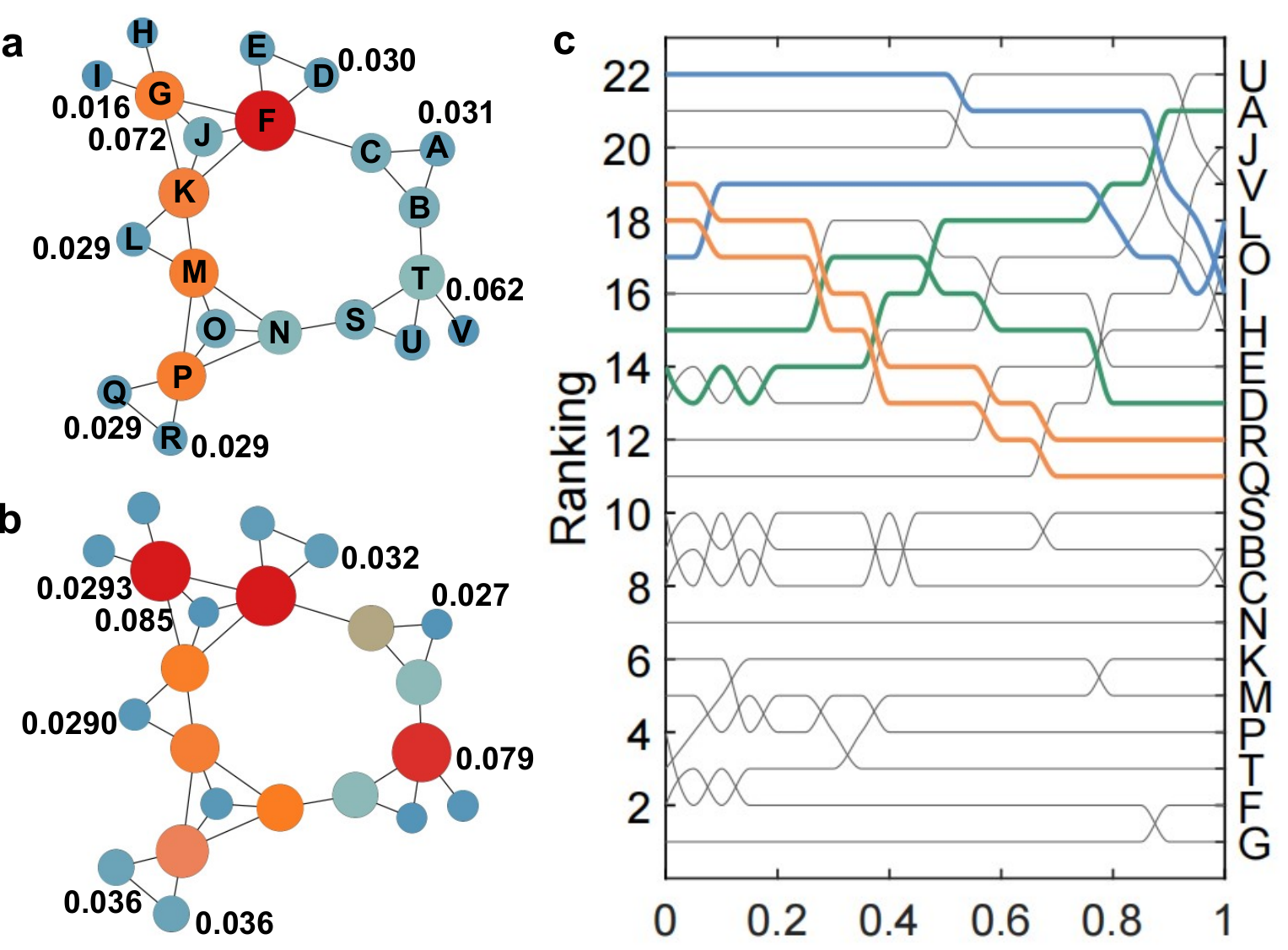}}
\caption{\textbf{Node ranking scores of a toy network and their variation with $s$.} Subfigures \textbf{a} and \textbf{b} display the node rankings under HoRW with $s=0$ and $s=1$, respectively. \textbf{c} displays the node ranking trends with different tuning parameters from pairwise (s=0) to higher-order (s=1). The sizes and colors of nodes are proportional to their importance.
\label{fig3}}
\end{figure}

We construct a toy network to scrutinize the variation tendency of the proposed Higher-order Augmented Random Walk (HoRW) on node ranking as $s$ changes, as depicted in Fig. \ref{fig3}. 
We pick three pairs of nodes to analyze the effectiveness of HoRW below. Additionally, we tested the effectiveness of the HoRW on several real-world networks.

\unskip

\textbf{Comparing Nodes $Q$ and $R$}. 
Nodes $Q$ and $R$ exhibit equal importance according to HoRW (s=0), as they are both constituents of the same triangle and exhibit symmetry in the network topology.
However, as we transfer from HoRW (s=0) method to HoRW (s=1), the importance of these two nodes progressively increases.
This is because, initially, they would be subject to a localization effect where the more important node $P$ would absorb the importance of $Q$ and $R$.
In our approach, however, the redistribution of information within the higher-order structures mitigates this effect, leading to an increase in the importance of these nodes.

\textbf{Comparing Nodes $A$ and $D$}. 
In HoRW (s=0), the most important node $F$ absorbs the importance of neighboring nodes, resulting in localization. Consequently, node $D$ is less important than $A$.
However, node $A$ and node $D$ belong to different triangles in the network, and the triangle containing $D$ is more important than the one containing $A$ on average. 
Considering higher-order structures alleviates the localization issue, and node $A$ gains more when the information is redistributed within the higher-order structures.
As a result, the distribution of importance among nodes diverges when higher-order interaction is emphasized, such as when $s=1$, culminating in the importance of node $D$ surpassing that of node $A$.


\textbf{Comparing Nodes $I$ and $L$}. 
Under HoRW (s=0), the importance score of node $I$ is lower than that of node $L$ as the most important structure, i.e., simplex $\left[F, G, J, K\right]$, and the most important node in it absorbs the importance of peripheral nodes $I$.
However, under the HoRW with $s=1$, the opposite is true, i.e., node $I$ is more important than node $L$. This is because $I$ and the most important node $G$ are located in the same simplex, and $I$ is adjacent to the most important structure - simplex $\left[F, G, J, K\right]$.


\subsection{Convergence of HoRW} 
%
Demonstrating the convergence of the HoRW \add{holds paramount significance} \del{is important}, as eliminating the necessity for an auxiliary convergence step in the algorithm would confer a significant advantage in scaling the model to large-scale networks \cite{lu2011leaders}.

We conduct convergence analysis on a connected network with diameter $l$. \add{First,} we show that the topologies of the networks corresponding to \add{the} matrices $C$ and $W$ are identical.
If $C_{ij}>0$, then nodes $i$ and $j$ are directly connected, and there exists at least one simplex containing both $i$ and $j$. We can find a path from $i$ to $\alpha$ and $\alpha$ to $j$, and therefore, $W_{ij}>0$. 
Similarly, if $W_{ij}>0$, we can find at least one simplex containing both $i$ and $j$, indicating that nodes $i$ and $j$ are directly connected in the original graph, i.e., $C_{ij}>0$. 
We can conclude that the elements in $C^l$ and $W^l$ are all positive, i.e., $C$ and $W$ are primitive since the network topology corresponding to the matrices $C$ and $W$ is the same and the network diameter is $l$.

The matrix $W^{\top}$, which is row-normalized, has obviously an eigenvalue $1$ with eigenvector $\frac{1}{N}$, and thus $1$ is also an eigenvalue of $W$. We will prove that the algebraic multiplicity of this eigenvalue is 1 by contradiction. Suppose \add{that the matrix} $W^{\top}$ has another eigenvector $\boldsymbol{x}$ whose entries are not identical and the largest one is $x_i$. The assumption of eigenvectors with heterogeneous terms leads to the contradiction that
\begin{equation}
     \boldsymbol{x}  = \left(W^l\right)^{\top} \boldsymbol{x} \Rightarrow x_i  = \sum_j w_{ij}^{\prime} x_j <\sum_j w_{ij}^{\prime} x_i = x_i.
\end{equation}
Here, $w_{ij}^{\prime} $ denotes the element of the $i$-th row and $j$-column row of matrix $(W^l)^{\top}$. This contradiction indicates that $(W^l)^{\top}$, and thus $W$, has a unique eigenvector associated with eigenvalue 1, i.e., a unique steady state. 

\add{Similarly} \del{Likewise}, we can prove that $C$ possesses a unique steady state, and so does the augmented transition matrix $\widetilde{M}$.

After analyzing the principles of node ranking under the toy example and demonstrating the convergence of the proposed model, we proceed to assess the effectiveness of the HoRW method on some real-world \del{complex} networks.

\add{
\subsection{Complexity analysis}
}

\add{
The complexity of the HoRW algorithm is similar to that of metrics based on random walk methodologies.
To complement these insights, we present comprehensive records of the running time, meticulously reported in Table S8 in the Supplementary Information.
These results underscore the competitiveness of HoRW in terms of computational complexity.
}

\add{
Separately, it is necessary to consider the one-time preprocessing phase for finding simplices.
On a more specific note, within a graph consisting of $n$ vertices and $m$ edges, the number of $\ell$-simplices is bounded by $O\left(n^{\ell-1}\right)$. Their enumeration can be done in $O\left( a\left(G\right)^{\ell-3} m\right)$ \cite{chiba1985arboricity}. Here, $a\left(G\right)$ is the arboricity of the graph $G$ - a measure of graph sparsity, and it is shown to be no more than $O(m^{1/2})$. Therefore, all $\ell$-simplices can be enumerated in $O\left( n^{\ell-3} m \right)$ as $m \leq n^2$.
}

\section{Experiments and Results}
\label{sec:exp}

In this section, we first briefly describe four empirical networks that are employed in experiments. Then we evaluate the effectiveness of HoRW in two extensively studied dynamical processes, namely epidemic spreading and network dismantling, followed by a node ranking resolution analysis.

\subsection{Data Description}
\label{sectionData} 
To evaluate the performance of HoRW, we conduct experiments on four real-world datasets from disparate fields:
\begin{itemize}
    \item \textbf{PolBooks} \cite{polbooks}: the network of books concerning US politics sold by the online bookseller Amazon.com (Valdis Krebs);
    \item \textbf{USAir} \cite{colizza2007reaction}: the US air transportation network acquired by considering the 500 airports with the highest traffic volume;
    \item \textbf{Grid} \cite{watts1998collective}: the Western States Power Grid network of the United States, wherein nodes correspond to the transformer or power relay points and an edge exists between two nodes if a power line runs between them;
    \item \textbf{LastFM} \cite{Lastfm2020feather}: the social network derived from LastFM, a music streaming platform. The nodes represent individual users located in Asian countries, while the edges denote mutual follower relationships between them.
\end{itemize}

The topological features of these networks are summarized in Table \ref{DatabaseTable}. These datasets exhibit varied topological features, allowing us to assess the adaptability and robustness of our proposed method across diverse network structures.

\begin{table}[!ht]
\centering
\caption{Basic topological features of the four real-world networks considered in this work.}
\label{DatabaseTable}
\resizebox{0.59 \textwidth}{!}{
\begin{tabular}{cccccccccccc}
\toprule
Network &$N$ &$M$ &$\left \langle k \right \rangle$ &$\left \langle k^2 \right \rangle$ &$\mathcal{C}$\\
\midrule
PolBooks    &105    &441     &8.40     &100.25       &0.49\\
USAir    &500  &2980  &11.92    &641.12      &0.62\\
Grid         &4941  &6594  &2.67  &10.33    &0.08\\
LastFM      &7624  &27806  &7.29  &185.44   &0.22\\
\bottomrule
\label{tab1}
\end{tabular}}
\begin{tablenotes}
    \item Here, $N$ and $M$ denote the number of nodes and \add{edges} \del{links} in the network, $\left \langle k \right \rangle$ and $\left \langle k^2 \right \rangle$ are the mean degree and the mean square degree, and $\mathcal{C}$ denotes the clustering coefficient.
\end{tablenotes}
\end{table}

\subsection{The Vital Nodes in the Spreading Process}
\label{section3}

In the prior experiment, we observed that accounting for higher-order structures would alter the identification of top nodes.
To further substantiate the effectiveness of the proposed approach in information and disease spreading, we evaluate the performance of HoRW in epidemic contagion experiments and compare it (with $s=0$, $s=0.5$, $s=1$) with five \add{traditional} centrality measures (namely, PageRank, Degree, Betweenness, Eigenvector, and Coreness) \add{and six new centrality measures (namely, Gravity, Leverage, Two-way random walk betweenness centrality, Quasi-Laplacian, Clustered-local-degree, and $k$th Laplacian-energy)}. 
Both the SIR \cite{lloyd2001viruses} and HSIR \cite{iacopini2019simplicial} models are employed to simulate the contagion process on four real-world networks: PolBooks, USAir, Grid, and LastFM.

We initiate the process with the top 5\% of nodes, ranked by different centrality methods, in the infected state, while the remaining nodes are in the susceptible state.
We also conduct experiments with the top 1\% and top 10\% of nodes, see \add{Supplementary Information Section 4} for more results.
Then, those susceptible nodes have a probability of $\beta$ of being infected by each of their infected neighbors, and those already infected nodes have the probability of $\gamma$ of recovering. The iteration proceeds until a steady state is reached, which means the number of recovered nodes no longer increases in subsequent steps.

\textbf{The Standard SIR Contagion Model Case.} 
In our experiments, we fix the recovery probability $\gamma$ as 1 and evaluate our proposed methodology alongside multiple existing techniques under varying degrees of infection rates.
Specifically, we experiment with three cases: $\beta = \beta_c$, $\beta = 2  \beta_c$ and $\beta = 4 \beta_c$ \add{(see Supplementary Information Section 4)}, with each experiment repeated 100 times.
Here, the spreading threshold $\beta_c$ \cite{lloyd2001viruses} is employed as the benchmark, and mathematically, it is defined as:
\begin{equation}
\beta_c=\frac{\left \langle k \right \rangle}{\left \langle k^2 \right \rangle-\left \langle k \right \rangle},
\end{equation}
where $\left \langle k \right \rangle$ and $\left \langle k^2 \right \rangle$ denote the average degree and the average square degree, respectively.


\begin{figure*}[!ht]
\centerline{\includegraphics[width=\linewidth]{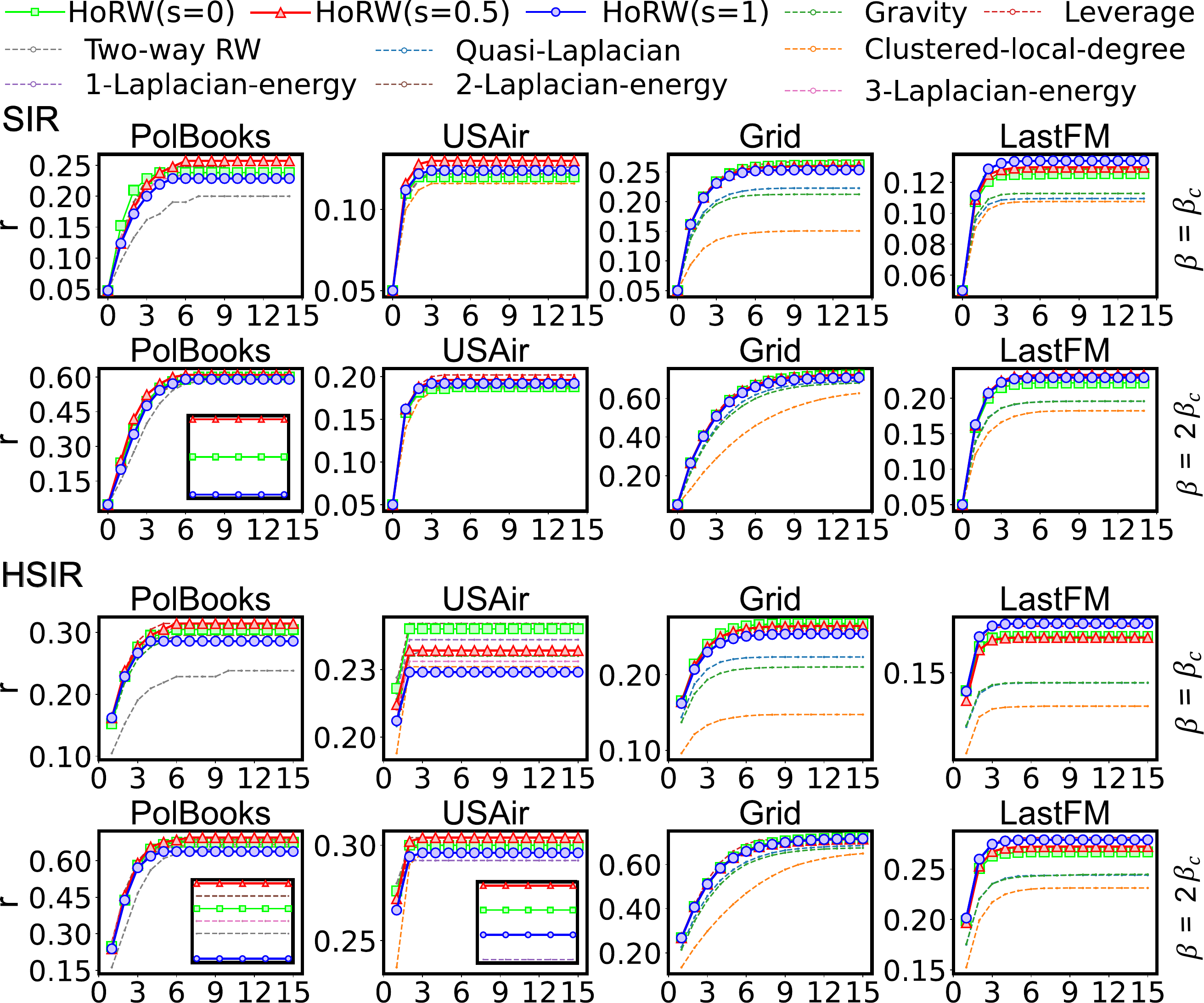}}
\caption{\textbf{ SIR spreading \add{under 5\% initial infected nodes} on four real-world networks.} The x-axis denotes the spreading steps, and the y-axis indicates the infection rate ($r$) in the spreading. 
The upper two panels depict epidemic spreading in classical SIR dynamics under $\beta = \beta_c$ and $\beta = 2 \beta_c$. And the bottom two panels depict epidemic spreading in classical HSIR dynamics under $\beta = \beta_c$, $\beta = 2 \beta_c$ and $\beta_2 = 0.8 \beta$.
}
\label{fig_low}
\end{figure*}   
\unskip

\begin{figure*}[!ht]
\centerline{\includegraphics[width=\linewidth]{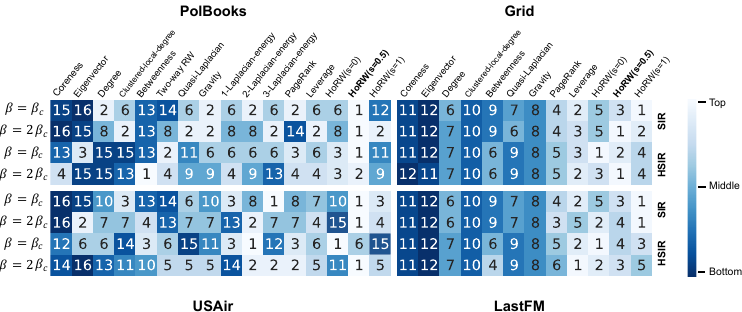}}
\caption{ \textbf{Ranking of SIR and HSIR spreading range on four real-world networks \add{under 5\% initial infected nodes}.} The color indicates the ranking of the final contagion range, with brighter indicating a greater contagion scope.
}
\label{fig: heatmap}
\end{figure*}

%
%
The upper two panels in Fig. \ref{fig_low} shows the results of SIR spreading experiments on four real-world networks \add{under several indexes}. The findings reveal that, among all these methods, the mixed method of HoRW (s=0.5) achieves a faster spread speed and wider range because it takes into account both pairwise and higher-order structures. 
The results vary across methods owing to their distinct emphasis on network properties.
To provide greater clarity, we visualize the ranking of infection rate in the steady state under six methods in Fig. \ref{fig: heatmap}. 
This figure confirms that our method achieves wider dissemination than other methods.

\unskip

\textbf{The HSIR Contagion Model Case.} 
In order to verify our method is also effective in higher-order tasks, we conduct HSIR simulation experiments to verify the effectiveness of the proposed model.
In this experiment, we consider simplices up to order 3 and perform HSIR simulations with three sets of parameters: $\beta = \beta_c$, $\beta = 2 \beta_c$, $\beta = 4 \beta_c$ \add{(see Supplementary Information Section 4)}, along with parameters $\beta_2 = 0.8 \beta$ for all $\beta$.
The efficacy of the epidemic on the four datasets is depicted in \add{bottom two panels in Fig. \ref{fig_low}}. 
It can be observed that the HoRW model exhibits better performance in identifying critical nodes under the higher-order epidemic spreading model.
We also visualize the ranking of infection scope in the steady state under six methods in Fig. \ref{fig_low}, which confirmed that our method achieves a wider spreading range than other methods.



Furthermore, the comparison between \add{the upper and bottom panels in} Fig. \ref{fig_low} suggests that the top nodes identified by HoRW perform better than those identified by other metrics, particularly in HSIR experiments.
These results suggest that the HoRW model outperforms not only pairwise tasks but also higher-order ones.

\subsection{The Vital Nodes in the Network Dismantling Problem}
\label{section4}

To further assess the collective influence of the vital nodes identified by HoRW to maintain the network's connectivity, we applied our algorithms to solve the network dismantling problem \cite{feng2023generalized}. The network dismantling problem aims to find an optimal set of nodes whose removal or deactivation can substantially diminish network connectivity, ultimately leading to the collapse of the network. 
Here we focus on the number of nodes that should be removed from a network to decrease the size of its giant connected component (GCC). The goal of our experiments is to make the GCC less than or equal to $1\%$ of the original size. 
We would prefer to eliminate fewer nodes to achieve this goal.
We compare our approach with some classical node ranking algorithms that are commonly used in network dismantling tasks \cite{wandelt2018comparative,pei2020influencer}, namely Betweenness \cite{freeman1977set}, Eigenvector centrality \cite{bonacich1987power}, Coreness \cite{kitsak2010identification}, and CoreHD \cite{zdeborova2016fast} \add{and six new centrality measures (namely, Gravity, Leverage, Two-way random walk betweenness centrality, Quasi-Laplacian, Clustered-local-degree and $k$th Laplacian-energy).} 
%
%
In the experiments, nodes will be removed one by one from real-world networks according to the descending order of their importance. The removal procedure will stop when the size of the GCC is less than or equal to $1\%$ of the original size.
After the removal process, there is a reinsertion operation \cite{Morone2015Nature,Ren2019Generalized}. In the case where the GCC size does not exceed $1\%$ of the original size, all the unnecessary removed nodes will be reinserted into networks again.
The proportion of removed nodes is shown in Table \ref{tab:dismantling}. \add{We also compare our HoRW with six new centrality measures in dismantling tasks, see Supplementary Information for more results.}

\begin{table}[!ht]
\centering
\caption{The proportion of removed nodes of different methods.}
\begin{tabular}{ccccc}
\toprule\toprule
Methods                & PolBooks & USAir    & Grid     & LastFM   \\  
\midrule 
CoreHD                    & 0.620  & 0.202   & 0.110  & 0.208 \\
PageRank                  & 0.620  & 0.214   & \R{0.060}  & 0.169 \\
Betweenness               & 0.670  & 0.198   & 0.062  & 0.172 \\
Degree                    & 0.670  & 0.226   & 0.159  & 0.172 \\
Coreness                  & 0.710  & 0.250   & 0.067  & 0.174  \\
Eigenvector               & 0.710  & 0.378   & 0.068  & 0.193\\                 
\midrule
Gravity                & 0.714    & 0.352    & 0.063    & 0.175    \\  
Leverage               & 0.628    & 0.212    & 0.061    & \R{0.168}    \\   
Two-way RW             & 0.724    & 0.194    & N/A      & N/A         \\
Quasi-Laplacian        & 0.714    & 0.356    & 0.063    & 0.193    \\   
Clustered-local-degree & 0.714    & 0.354    & 0.070    & 0.190    \\   
1-Laplacian-energy     & 0.714    & 0.250    & N/A      & N/A         \\   
2-Laplacian-energy     & 0.714    & 0.356    & N/A      & N/A         \\   
3-Laplacian-energy     & 0.714    & 0.362    & N/A      & N/A         \\   
\R{HoRW} & \R{0.610} & \R{0.186} & \R{0.060} & \R{0.168} \\ 
\bottomrule\bottomrule
\end{tabular}
\begin{tablenotes}
\item \add{The number of each element denotes the proportion of nodes that need to be removed until the GCC size is 0.01. The node importance calculated by Two-way RW and k-Laplacian-energy indicators on Grid and LastFM datasets cannot be obtained because of memory and time, these elements are filled with N/A. The best results are colored in red.}
\end{tablenotes}
\label{tab:dismantling}
\end{table}

The findings reveal that our proposed HoRW algorithm demonstrates remarkable performance in tackling the network dismantling problem, outperforming comparing methods. 
This highlights our method's strong capability to identify not only the most influential nodes but also a critical set of nodes whose removal can significantly disrupt the network structure.

\begin{figure*}[!t] 
\centerline{\includegraphics[width=\linewidth]{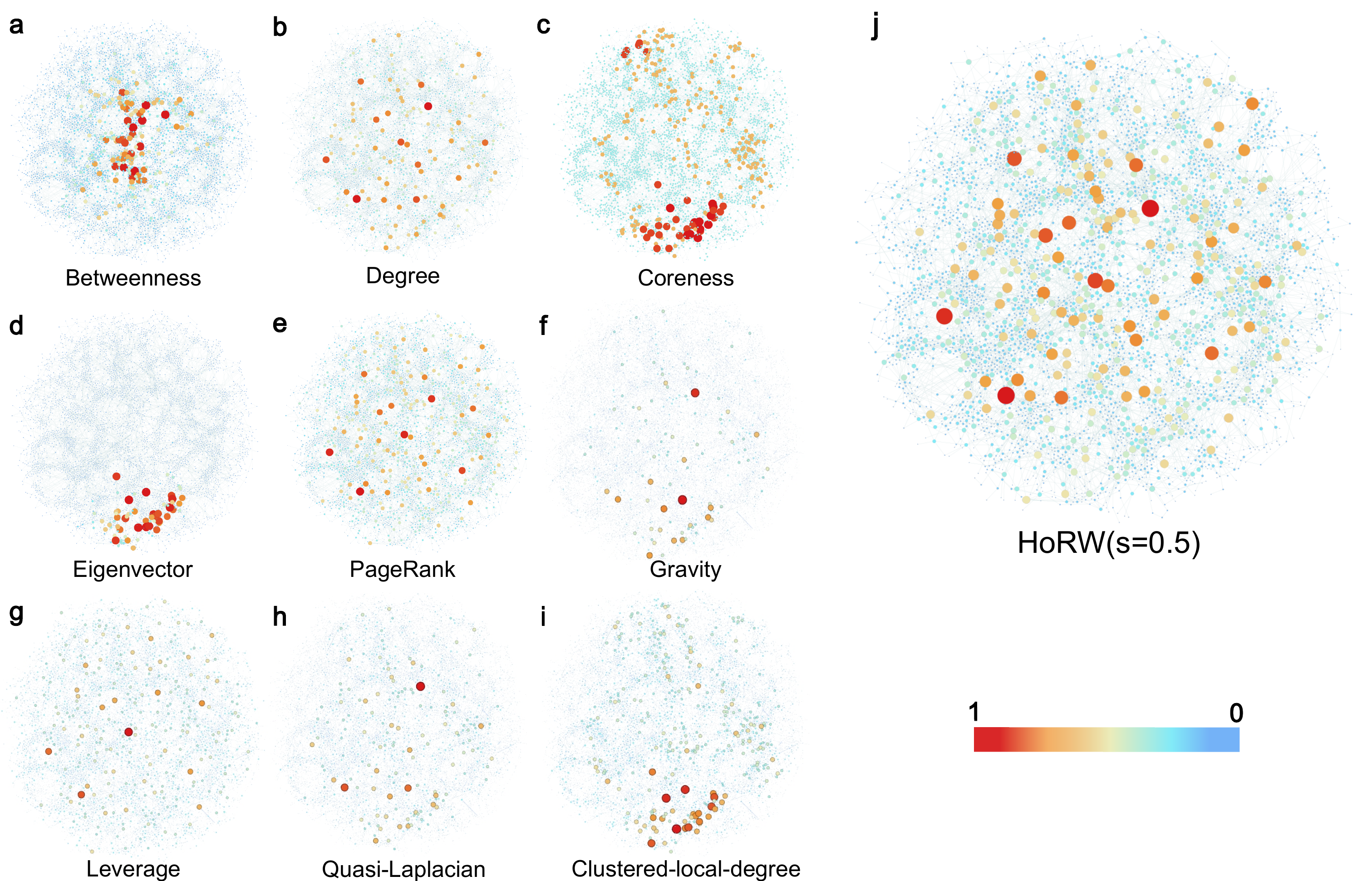}}
\caption{ \textbf{Visualization of the rankings of nodes in the Grid network.} The Grid network is visualized in several metrics from \textbf{a} to \textbf{i}. \textbf{j} denotes the visualization of the network under HoRW (s=0.5). 
The node colors are proportional to their min-max normalized values, and the position of each node in each plot remains fixed. 
\label{fig: visualization}}
\end{figure*}   
\unskip

\subsection{The Node Ranking Resolution}
\label{subsec: resolution}

The aim of this section is to clarify the advantages of the proposed HoRW method in terms of node ranking resolution, i.e., the value gap between two arbitrary sequential nodes after sorting according to centrality scores.

\add{Since the results ranking of propagation and dismantling dynamics are not exactly the same for various methods under different datasets.}
We conjecture that this results from different methods emphasizing distinct network factors,  thereby manifesting varying resolutions.
Hence, we visualize the node ranking of the Grid network with six different metrics in Fig. \ref{fig: visualization}. 
Here, min-max normalization of node scores is applied to compare different metrics, and it follows that:
\begin{equation}
x^*=\frac{x-x_{min}}{x_{max}-x_{min}} .
\end{equation}

\add{Subfig. \ref{fig: visualization} \textbf{a}, \textbf{d} and \textbf{i} show that the Betweenness, Eigenvector and Cluster-local-degree centrality methods exhibit a localization trait, wherein a few nodes are substantially prominent and tightly interconnected, while the remaining non-top nodes can barely be distinguished. This phenomenon is also known as the ``rich club".
Similarly, coreness centrality (subfig. \ref{fig: visualization} \textbf{c}) exhibits a localization property, but at the same time, it assigns significant and uniform importance to middle nodes. 
While the subfig. \ref{fig: visualization} \textbf{b}, \textbf{e} and \textbf{g} show that the important nodes are scattered, but the number of middle nodes identified is small, and there are still many bottom nodes that are difficult to distinguish.
The remaining subfig. \ref{fig: visualization} \textbf{f} and \textbf{h} (Gravity and Quasi-Laplacian) can only distinguish fewer vital nodes and the importance scores of the remaining nodes become small and indistinguishable.}

On the contrary, it can observe that nodes are extensively distributed throughout the entire network when scored by HoRW, and the distribution of middle nodes is notably more expansive compared to other metrics.

Empirical approaches to ranking nodes generally excel at identifying top nodes, yet lack sufficient resolution for the remaining nodes. 
As a consequence, a considerable number of nodes end up with identical or very similar scores, which makes it challenging to tell them apart.
Notwithstanding, these non-top nodes, particularly the middle nodes, exert significant influence due to their wide distribution within networks, despite not being as crucial as the top ones. 
A higher resolution ranking method would facilitate distinguishing each node in the network according to its importance, rather than solely identifying the top nodes.
In light of the disparities between the distributions of various metrics observed in Fig. \ref{fig: visualization}, we endeavor to quantitatively assess the resolution level of the distribution of HoRW and attain a higher resolution by adjusting $s$.
Specifically, the tuning parameter $s$ enables multiscale node ranking in accordance with the strength of higher-order effects.
However, owing to the heterogeneity of network structures, $s$ corresponding to optimal resolution differs for each network.
Thus, identifying the optimal tuning parameter $s$ for each network is essential to achieve a higher resolution for HoRW.

In an ideal scenario, each node in the network would be distinguishable by its importance score, requiring that node importance scores decrease uniformly after being ordered from largest to smallest.
We thus select $y=-x+1$ as the optimal importance sequence, that is, the benchmark sequence, where greater proximity to this function signifies increased distinguishability among nodes.
Besides, we leverage cosine similarity \cite{tan2016introduction} to quantify the proximity of different metrics to the benchmark, defined as follows:
\begin{equation}
        Cos(A,B)
        =\frac{\sum_{i=1}^n(A_i \times B_i)}{\sqrt{\sum_{i=1}^n (A_i)^2} \times \sqrt{\sum_{i=1}^n (B_i)^2}} 
        =\frac{A \cdot B}{|A| \times |B|}.
\end{equation}
To ascertain the optimal $s$ for HoRW, we compute the \add{cosine similarity} between the importance sequence and the benchmark $y=-x+1$ with $s$ ranging from $0$ to $1$ in the step of $0.01$, and the optimal tuning parameter is found to be $s=0.57$ for the Grid network (\add{the selection process is described in Section 2 of Supplementary Information}).

%

\begin{figure*}[!ht]
\centerline{\includegraphics[width=\linewidth]{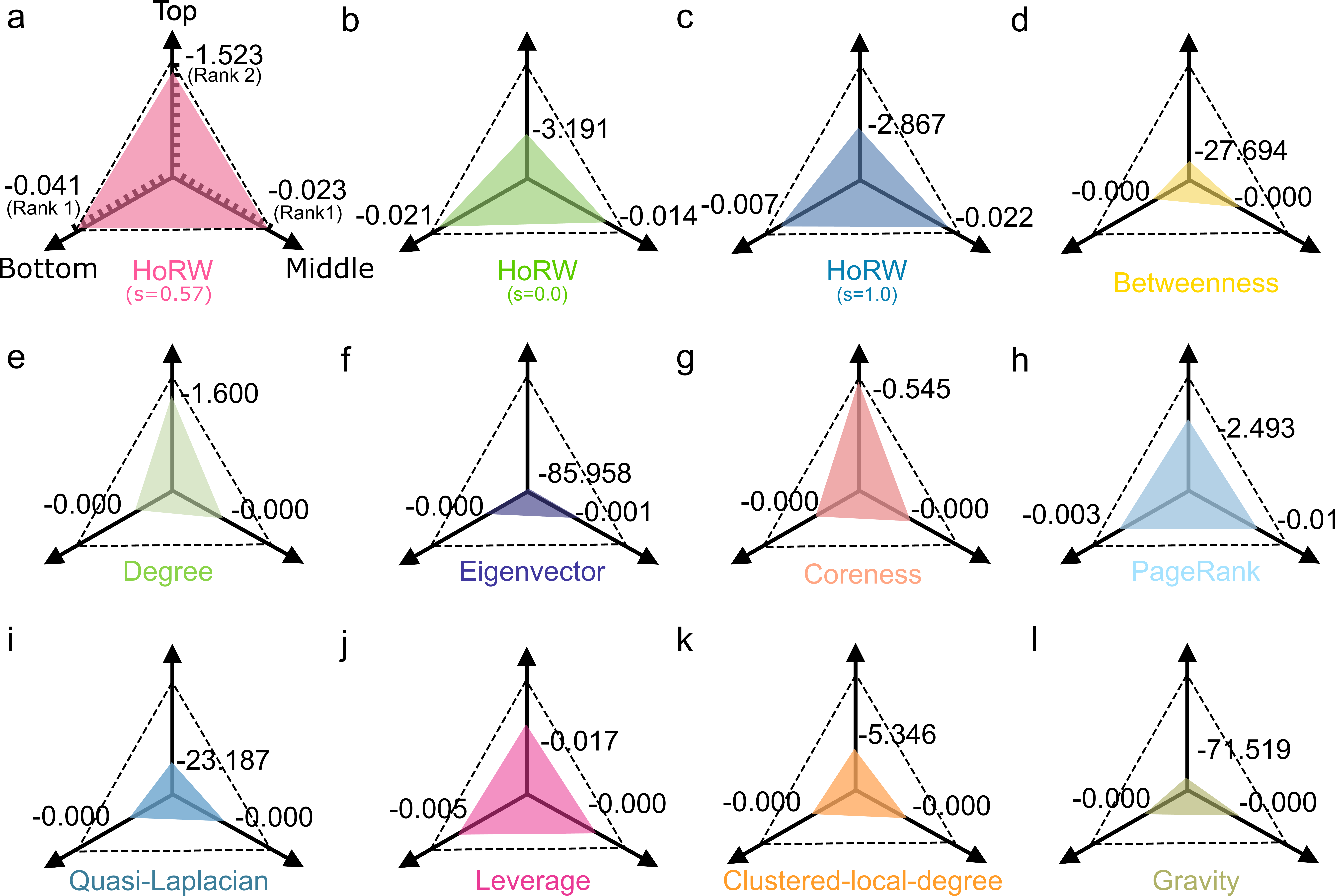}}
\caption{ \textbf{Comparison of the distances between each method and the benchmark\add{s}.} 
\textbf{a} illustrates the ranking under our proposed HoRW (s=0.57), in which the x,y and z axes represent the bottom, middle and top 50 nodes resolution rank. The dot triangle in (a) denotes the optimal result while the origin represents the worst result in these methods. 
\textbf{b} and \textbf{c} represent the result of our proposed HoRW under $s=0.0$ and $s=1.0$. And \textbf{d}-\textbf{i} denote the resolution result under other methods.
\label{fig:3D}}
\end{figure*}

To streamline the evaluation process, we partition the ranking into three discrete categories: top, middle, and bottom nodes.
Specifically, we compute the slope for the top 50, middle 50, and bottom 50 nodes (1\% of the Grid network's size), respectively.
A slope proximity to $-1$ signifies that it is closer to the baseline sequence $y=-x+1$, indicating a higher resolution.
We then present the top, middle, and bottom resolutions of each method in 3-dimension space, with the x, y, and z axes representing the slope from the bottom, middle, and top perspectives, as shown in Fig. \ref{fig:3D}, where the dotted triangles are the benchmark (optimal results).
Additionally, Table S8 in SI quantitatively compares \del{our} HoRW \del{method} with other methods from these three perspectives.


\unskip

From Fig. \ref{fig:3D} and Table S8, we can draw that the Eigenvector centrality and other methods focus more on finding important nodes but sacrifice resolution.
Conversely, HoRW demonstrates closer proximity to the benchmark than other methods. We find that HoRW (s=0.57) gets the optimal performance of resolution in the middle and bottom perspective in the Grid dataset. While in distinguishing top nodes, our method can also get a decent result.
These findings indicate that the proposed method is more efficient in distinguishing middle and bottom nodes while maintaining a balance across all three perspectives, demonstrating its better performance with respect to ranking resolution.

To sum up, our experimental findings demonstrate that HoRW outperforms other methods in terms of node ranking resolution, explaining its advantage in epidemic spreading and network dismantling. Moreover, we have discovered that it is a high-resolution ranking algorithm that can be fine-tuned by selecting an appropriate tuning parameter $s$.

\section{Conclusion}
\label{sec:conclusion}

Vital node identification is a fundamental issue in network science since empirical networks possess heterogeneous connections. 
Existing methods typically operate on pairwise networks without accounting for higher-order structures.
In this paper, we introduce a multi-order graph by integrating the higher-order bipartite graph, which captures higher-order features, with the classical pairwise graph. 
Furthermore, inspired by random walk dynamics and higher-order topology theory, we propose a Higher-order Augmented Random Walk (HoRW) model to identify vital nodes in networks. 
HoRW enables multiscale node ranking by adjusting a tuning parameter $s$ and considers not only binary connections between nodes but also higher-order interactions involving multiple nodes, making it more comprehensive in capturing the network structure and dynamic features.

To demonstrate the advantages of HoRW, we examine two application scenarios: the spreading process and the network dismantling problem.
In the spreading models, seed nodes are crucial to the final spreading scope. 
Specifically, if the selected seed nodes are influential and widely distributed, then the contagion scope is prone to be larger.
Besides, middle nodes can further accelerate contagion propagation.
As for network dismantling, identifying and attacking critical nodes can effectively dismantle the entire network, and effectively pinpointing middle nodes can help achieve this goal when the target size of the GCC is small.
Empirical experiments demonstrate that the proposed method performs well in both the spreading process and network dismantling tasks.
Moreover, the high resolution of the proposed method makes nodes more differentiated in terms of the centrality scores, rather than just identifying the most vital nodes, 
giving its advantages in epidemic spreading and network dismantling.

In conclusion, our model is effective in many tasks and highly scalable. 
We promise that HoRW can provide a fresh perspective on vital node identification and open up new avenues for exploring the recognition of higher-order structures in networks. 
Beyond node identification tasks, it promises to be applied to more complicated tasks such as link prediction and high-influential community mining in social networks.

\section*{CRediT authorship contribution statement}
\textbf{Yujie Zeng}: Data curation, Methodology, Software, Formal analysis, Visualization, Writing - original draft, Writing - review \& editing.
\textbf{Yiming Huang}: Investigation, Formal analysis, Visualization, Writing - original draft, Writing - review \& editing.
\textbf{Xiao-Long Ren}: Conceptualization, Data curation, Methodology, Formal analysis, Writing - review \& editing, Supervision, Funding acquisition.
\textbf{Linyuan L{\"u}}: Conceptualization, Methodology, Writing - original draft, Writing - review \& editing, Supervision, Resources, Funding acquisition.

\section*{Declaration of Competing Interest}
The authors declare that they have no known competing financial interests or personal relationships that could have appeared to influence the work reported in this paper. 

\section*{Acknowledgements}
The authors are grateful for the support from the STI 2030--Major Projects (2022ZD0211400), the National Natural Science Foundation of China (Grant No. T2293771), the China Postdoctoral Science Foundation (2022M710620), the Sichuan Science and Technology Program (2023NSFSC1353), the Project of Huzhou Science and Technology Bureau (2021YZ12), the UESTCYDRI research start-up (U032200117), and the XPLORER PRIZE.

\section*{Code and Data Availability}
The authors declare that the code and data supporting the findings of this study will be available after this paper is published at the following GitHub repository: \href{https://github.com/sssleverlily/HoRW}{https://github.com/sssleverlily/HoRW}.

\bibliographystyle{elsarticle-num} 
\bibliography{ref}




\end{document}